\documentclass[11pt]{article}

\usepackage{amsmath}
\usepackage{amssymb}
\usepackage{epsfig}

\newcommand{\bea}{\begin{eqnarray*}}
\newcommand{\eea}{\end{eqnarray*}}

\begin{document}


\begin{titlepage}

\renewcommand{\thefootnote}{\alph{footnote}}
\vspace*{-3.cm}
\begin{flushright}

\end{flushright}

\vspace*{0.5cm}

\renewcommand{\thefootnote}{\fnsymbol{footnote}}
\setcounter{footnote}{-1}

{\begin{center} {\Large\bf The Noncommutative Quadratic Stark
Effect For The H-Atom}

\end{center}}
\renewcommand{\thefootnote}{\alph{footnote}}
\vspace*{.8cm} {\begin{center} {\large{\sc
                Noureddine~Chair$^{a,b}$, Mohammad.A.~Dalabeeh$^{a}$
                }}
\end{center}}
\vspace*{0cm} {\it
\begin{center}
 $^a$Physics Department,
Al al-bayt University, Mafraque, Jordan

Email: n.chair@rocketmail.com\\
\hspace{19mm}nchair@alalbayt.aabu.edu.jo
\end{center} }

\vspace*{0cm} {\it
\begin{center}
 $^b$The Abdus Salam International Centre For Theoretical Physics,
 1-34014 Trieste,Italy,

Email: chairn@ictp.trieste.it
\end{center} }

\vspace*{0cm}

\vspace*{1.5cm}

{\Large \bf
\begin{center} Abstract \end{center} }
Using both the second order correction of perturbation theory and
the exact computation due to Dalgarno-Lewis, we compute the second
order noncommutative Stark effect,i.e., shifts in the ground state
energy of the hydrogen atom in the noncommutative space in an
external electric field. As a side result we also obtain a sum
rule for the mean oscillator strength. The energy shift at the
lowest order is quadratic in both the electric field and the
noncommutative parameter $\theta$. As a result of noncommutative
effects the total polarizability  of the ground state is no longer
diagonal.

\vspace*{.5cm}

\end{titlepage}

\newpage

\renewcommand{\thefootnote}{\arabic{footnote}}
\setcounter{footnote}{0}


%

\section{Introduction \label{sec:SEC-intro}}
When an atom is subjected to an external electric field in a given
direction the energy levels of the atom shift; this is the Stark
effect (see e.g. Bethe and Salpeter 1957). If the electric field
is weak the shift in the energy level is linear in the electric
field, by parity arguments there will be shift only between states
of opposite parity so the linear shift depends on the the presence
of degenerate states. Therefore the ground state first order
effect vanishes and hence the energy shifts start at the quadratic
order, the quadratic Stark effect. In both the linear and the
quadratic Stark effect one uses the ordinary perturbation theory.
If the external field is strong, perturbation theory is not
justified and one may use the WKB method (see e.g Bekenstein and
Krieger (1969)). The combined Coulomb plus the external potential
felt by an electron implies that there are no absolutely stable
bound states, and there is a possibility of quantum tunnelling. If
the field is increased enough, then the states become unbound,
leading to the  ionization (see e.g Yamabi et al (1977)).

Here we are interested in studying the Stark effect in
noncommutative quantum mechanics of the H-atom in the ground state
for which one can safely use ordinary perturbation theory. In this
direction, the linear Stark effect has been studied by (Chaichian
et al (2001)) in which they showed that there is no Stark shift
linear in the electric field induced by the noncommutativity of
the coordinates and in particular the ground state remains
unchanged, in the linear Stark effect. Like the ordinary Stark
effect of the hydrogen atom, in order to obtain correction for
ground state energy one has to go to  the second order in
perturbation theory. In this paper we compute the noncommutative
Quadratic Stark energy shift using two methods. The first one, is
to use the second order in perturbation theory that we just
mentioned,  which contains an infinite summation over the excited
states. The other method available is that of the Dalgarno-Lewis
method (Dalgarno and Lewis (1955)) which is exact and does not
contain any summations over the excited states. In doing so we
show that there is a noncommutative Quadratic Stark effect which
results in a lowering the ground state energy like the ordinary
quadratic Stark effect. Both of these energies contribute
additively to the polarization of the hydrogen atom  and as a
result of the noncommutativity of the polarization tensor is no
longer diagonal. When equating the two expressions for the
noncommutative quadratic Stark shift in energy obtained by the two
methods, we obtain the sum rule on the mean oscillator strength
for the Lyman series.

\section{The Noncommutative Quadratic Stark Effect For the Ground State of the H-atom}
Noncommutative quantum mechanics is defined through the following
commutation relations,
\begin{eqnarray}
\label{t0t01}
 \left[\mathbf{x}_i  ,\mathbf{x}_j\right]
&=&i\theta_{ij}\  \cr \left [\mathbf{x}_i ,\mathbf{p}_j\right] &=&
i\hbar\delta_{ij}\  \cr
 \left[\mathbf{p}_i,\mathbf{p}_j\right] &=&
0\,
\end{eqnarray}
where $\theta_{ij}$ is the noncommutative parameter and is of
dimension of $(length)^2$. It is convenient, however, for
computational reasons to change to the new coordinate system,
\begin{eqnarray}
\label{toto2}
x_i&=&\mathbf{x}_i~+\frac{1}{2\hbar}\theta_{ij}\mathbf{p}_j\ , \cr
p_i&=&\mathbf{p}_i,\,
\end{eqnarray}
where the new coordinates satisfy the usual canonical commutation
relations:
\begin{eqnarray}
\label{toto3}
 [ x_i , x_j ] = 0,~~~[ x_i , p_j ] = i\hbar\delta_{ij}~~~
[ p_i , p_j ] = 0.
\end{eqnarray}
When the electron in the H-atom is subjected to a uniform electric
field E in the positive z-direction, its potential energy is
$eE\mathbf{z}$, therefore using equation (\ref{toto2}) the total
Hamiltonian of this system that takes into account the
noncommutative coordinates is,
\begin{equation}
\label{toto4}
 H=\frac{p^2}{2m} -\frac{e^2}{r} +eEz
+\frac{e}{4\hbar}(\vec{\theta}\times \vec{p})\cdot\vec{E}.
\end{equation}

The term $eEz$ is the ordinary Stark potential whereas the term
$+\frac{e}{4\hbar}(\vec{\theta}\times \vec{p})\cdot \vec{E}$ in
the above equation corresponds to the noncommutative Stark
potential energy. It is well known that at the first order in
perturbation theory there is no shift in the ground state energy
of the H-atom  due to the perturbing term $eEz$. Similarly the
noncommutative Stark perturbing potential does not change the
ground state energy in the first order of perturbation theory
(Chaichian et al (2001)). Like the ordinary Stark effect one needs
to go to the second order in perturbation theory in order to find
the correction to the ground state energy, this the quadratic
Stark effect. Using the fact that $p_i=\frac{m}{i\hbar}[x_i,H_o]$,
where $H_{0}=\frac{p^2}{2m} -\frac{e^2}{r}$ the correction to the
ground state energy then is,

\begin{eqnarray}
\label{toto5} E_{1}^{2(NC)}&=&\sum_{n\neq
1,l,m}\frac{|\langle n,l,m| H^{(NC)} |1,0,0\rangle|^2}{E_{n}^0-E_{1}^0} \nonumber\\
&=& \sum_{n\neq 1,l,m} \frac{|\langle n,l,m|\frac{e}{4
i\hbar^2}(\vec{E}\times
\vec{\theta})_1([x_1,H_o]+(\vec{E}\times\vec{\theta})_2[x_2,H_o])
|1,0,0\rangle|^2}{E_n^0-E_1^0} ,
\end{eqnarray}
where we have used $({\vec{\theta}\times \vec{p}})\cdot
\vec{E}=({\vec{E}\times \vec{\theta}})\cdot \vec{p}$. The states
$|n,l,m\rangle$ form a complete set of eigenstates of the free
hamiltonian, $H_0|n,l,m\rangle=E_{n}^0|n,l,m\rangle$ and
$E_{n}^o=\frac{E_{1}^{0}}{n^2}$, $E_{1}^{0}$ is the ground state
energy. From the following equations,
\begin{eqnarray}
x&=&r\sqrt{\frac{2\pi}{3}}[\mathbf{Y}_{1}^{-1}(\theta,\phi)-\mathbf{Y}_{1}^{1}(\theta,\phi)]\nonumber\\
y&=&ir\sqrt{\frac{2\pi}{3}}[\mathbf{Y}_{1}^{-1}(\theta,\phi)+\mathbf{Y}_{1}^{+1}(\theta,\phi)],
\nonumber
\end{eqnarray}
we learn that the selection rules are, $ \Delta {m}=\pm 1$,
$\Delta{l}=1 $. Using the selection rules and doing the angular
integration, equation (\ref{toto5}) becomes,
\begin{equation}
\label{toto6} E_{1}^{2(NC)}=\frac{e^2 m^2
E_{1}^0}{16\hbar^4}|\vec{E}\times\vec{\theta}|^2~~
\frac{1}{3}\sum_{n\neq
1}\frac{n^2-1}{n^2}~~~|\int_{0}^{\infty}\mathbf{R}_{n1}r^{3}\mathbf{R}_{10}
dr|^2,
\end{equation}
where the radial integration is already computed and is given in
( Bethe and Salpeter 1957), the results is,
\begin{equation}
|\int_{0}^{\infty}\mathbf{R}_{n1}r^{3}\mathbf{R}_{10} dr|^2 =
\frac{2^8 n^7 (n-1)^{2n-5}} {(n+1)^{2n+5}}a^2, \nonumber
\end{equation}
where $a=\frac{\hbar^2}{me^2}=-\frac{e^2}{2E_1^0}$ is the Bohr
radius. Finally, the noncommutative quadratic Stark  effect for
the Hydrogen atom at the ground state reads
\begin{equation}
\label{toto7} E_{1}^{2(NC)}=-\frac{e^2
m}{32\hbar^2}|\vec{E}\times\vec{\theta}|^2~~\frac{1}{3}\sum_{n\neq
1}\frac{2^8 n^5 (n-1)^{2n-4}}{(n+1)^{2n+4}}.
\end{equation}
Therefore there is a contribution to the Stark effect from the
noncommutativity of  coordinates which were not present in the
first order in perturbation theory. Note also like the ordinary
Stark effect the noncommutative quadratic Stark effect also
results in a reduction of the the ground state energy. This means
that the contribution to the Stark effect would be the addition of
the usual quadratic Stark effect $E_{1}^{2(C)}=-\frac{9 a^{3}
}{4}|\vec{E}|^2$  and the one given by equation (\ref{toto7}).
There is no contribution from the crossed term of the full
potential,
\begin{equation}
V_{Stark}=eEz +\frac{e}{4\hbar}(\vec{\theta}\times
\vec{p})\cdot\vec{E}.\nonumber\\
\end{equation}
One can check this using the selection rules. In the next section
using the method due to (Dalgarno and Lewis (1955)) we will show
that the exact contribution to the noncommutative quadratic Stark
effect is $E_{1}^{2(NC)}=-\frac{e^2
m}{32\hbar^2}|\vec{E}\times\vec{\theta}|^2$
\section{The Exact Computation }
In this section we will briefly review the method due to (Dalgarno
and Lewis (1955)), then use it to carry out the computation for
the noncommutative quadratic Stark effect. In this method the goal
is to find the perturbed ground state $\psi_1^{1}$ which is
assumed to be related to the ground state in the form,
$\psi_1^{1}=F\psi_1$ where $F$ is a scalar function of the
variables that occur in the Hamiltonian, and then use the
relation,
$E_{1}^{2(NC)}=\langle\psi_1|H'|\psi_1^{1}\rangle-E_{1}^{1(NC)}\langle\psi_1|\psi_1^{1}\rangle=
\langle\psi_1|H'|\psi_1^{1}\rangle$ (the first order correction
$E_{1}^{1(NC)}=0$) to compute the second order correction, $H'$
being the noncommutative perturbing Hamiltonian and $\psi_1$ is
the ground state wave function for the hydrogen atom. The first
order equation in perturbation theory is given by,
\begin{equation}
\label{toto8}
 (H_{0}-E_{1})\psi_1^{1}+(H'-E_{1}^{1})\psi_1.
\end{equation}
Setting $\psi_1^{1}=F\psi_1$ where $F$ is a scalar function of the
variables that occur in the Hamiltonian, and define a reduced
potential by $V'=
H'-E_{1}^{1}=H'-\langle\psi_1|H'|\psi_1^{1}\rangle$ then the first
order perturbation equation it can be written as
\begin{equation}
\label{toto9} [ H_{0} , F ]\psi_1 +V'\psi_{1}=0,
\end{equation}
where $H_{0}=-\frac{\hbar^2}{2m}\nabla^{2}+u(r)$ is the free
Hamiltonian for the H-atom and $u(r)$ is the coulombic potential.
Because $u(r)$ and $F$ commute then one can show (e.g see Hmmeka
(1981)) that the function $F$ is the solution to the following
differential equation,

\begin{equation}
\label{toto10}
\frac{\hbar^2}{2m}\nabla^{2}F\psi_1-\frac{2}{a}\Omega F \psi_1=
\left(\frac{V'\psi_1}{\psi_1} \right),
\end{equation}
where $\Omega $ is the differential operator defined by,

\begin{equation}
\label{toto11}
 \Omega=\frac{x}{r}\frac{\partial}{\partial
x}+\frac{y}{r}\frac{\partial}{\partial
y}+\frac{z}{r}\frac{\partial}{\partial z}.
\end{equation}
The reduced potential in our case is the noncommutative Stark
potential as the first order correction $E_{1}^{1(NC)}=0$. By
acting by the operator $V_{Stark}^{(NC)}=
\frac{e}{4\hbar}(\vec{\theta}\times\vec{p})\cdot\vec{E}$ on the
ground state wave function, $\psi_1=(\pi{a}^3)^{-1/2}\exp{-(r/a)}$
the differential equation (\ref{toto10}) is then,
\begin{equation}
\label{toto12}
-\frac{\hbar^2}{2m}\nabla^{2}F-\frac{2}{a}\Omega F
= \frac{e}{4i}\left[(\vec{E}\times
\vec{\theta})_x\frac{x}{r}+(\vec{E}\times\vec{\theta})_y\frac{y}{r}
\right].
\end{equation}
From the following relations

\begin{eqnarray}
&&\nabla^{2}({x}{r})= \frac{4x}{r} ,~~~~~~~~~~~~~~~~~~~~~~~~~~~\Omega({x}{r})=2x \nonumber\\
&&\nabla^{2}(x)=0,~~~~~~~~~~~~~~~~~~~~~~~ \Omega(x)=\frac{x}{r} \nonumber\\
&&\nabla^{2}({y{}r})=\frac{4y}{r} ,~~~~~~~~~~~~~~~~~\Omega({y}{r})=\frac{4y}{r}\nonumber\\
&&\nabla^{2}(y)=0    ,~~~~~~~~~~~~~~~~~~~~~~~~~~\Omega(y)=1
\nonumber,
\end{eqnarray}
the solution to our differential equation would be of the form,
\begin{equation}
\label{toto13}
 F= \alpha x+\beta ({x}{r}) +\gamma y+\delta(yr),
\end{equation}
where $\alpha, \beta, \gamma,$ and $\delta$ are constants to be
found. Substituting this expression in equation, (\ref{toto12})
then we find,
\begin{equation}
\label{toto14}
F= \frac{em}{4i\hbar^2}\left[(\vec{E}\times
\vec{\theta})_x{x}+(\vec{E}\times\vec{\theta})_y {y} \right].
\end{equation}
Having found the expression for $F$, then the second correction to
the ground state energy will be given by an integration rather
rather than summation,
\begin{eqnarray}
\label{toto15}
E_{1}^{2(NC)}&=&\langle\psi_1|H'|F\psi_1\rangle \nonumber\\
&=& \frac{em}{4i\hbar^2}\langle\psi_1|\left[\vec{E}\times
\vec{\theta})_x{p_x}+(\vec{E}\times\vec{\theta})_y
{p_y}\right]\left[(\vec{E}\times
\vec{\theta})_x{x}+(\vec{E}\times\vec{\theta})_y
{y}\right]|\psi_1\rangle\ \nonumber\\
&=&-\frac{em}{16\hbar^2}\left[|\vec{E}\times\vec{\theta}|^2+
|(\vec{E}\times\vec{\theta})_x|^2\langle\psi_1|x\frac{\partial}{\partial
x}|\psi_1\rangle+
|(\vec{E}\times\vec{\theta})_y|^2\langle\psi_1|y\frac{\partial}{\partial
y}|\psi_1\rangle\right]\large.
\end{eqnarray}
Note that we have dropped the terms
$\langle\psi_1y\frac{\partial}{\partial x}\psi_1\rangle $ and
$\langle\psi_1yx\frac{\partial}{\partial y}\psi_1\rangle$ as they
vanish by orthogonality of the spherical harmonics. Writing both
$x$ and $y$ in the spherical harmonics then doing the angular and
radial integration we obtain the following exact result for the
noncommutative quadratic Stark effect,
\begin{equation}
\label{toto16}
E_{1}^{2(NC)}= -\frac{e^2
m}{32\hbar^2}|\vec{E}\times\vec{\theta}|^2 .
\end{equation}
Comparing this result with that obtained in the last section using
the second order perturbation theory given by equation
(\ref{toto7}) we conclude that ,
\begin{equation}
\label{toto17}
\frac{1}{3}\sum_{n\neq 1}\frac{2^8 n^5
(n-1)^{2n-4}}{(n+1)^{2n+4}}=1.
\end{equation}
Indeed this is the case, this sum is nothing but the sum rule of
the mean oscillator strength for the Lyman series ( Bethe and
Salpeter 1957) which is equal to 1 as one can check by setting
$l=0$ in equation (61.5) for the ground state ( Bethe and Salpeter
1957). Therefore, equating the exact computation for the energy
contribution in the noncommutative quadratic Stark effect with
that computed in the last section we derived the sum rule for the
mean oscillator strength. This quantity is a dimensionless
quantity and proportional to the frequency times the dipole
moment.

Adding the noncommutative quadratic Stark shift energy to the
ordinary Stark shift we have,
\begin{equation}
\label{toto18}
 E_{1}^{Stark}=-\frac{9 a^{3} }{4}|\vec{E}|^2
-\frac{e^2 m}{32\hbar^2}|\vec{E}\times\vec{\theta}|^2-\frac{3555
a^{7} }{64}|\vec{E}|^4+O(|\vec{E}|^6).
\end{equation}
This energy in terms of polarizability  is
\begin{equation}
\label{toto19}
 E_{1}^{Stark}=-\frac{1}{2}\varepsilon_{i j}E_{i}E_{j},
\end{equation}
where $\varepsilon_{i j}$ is the polarizability tensor. In general
$\varepsilon_{i j}$ is a function of external field, however, up
to the second order it is electric field independent. One can
easily read-off what is $ \varepsilon_{i j}$ from equation
(\ref{toto18}):
\begin{equation}
\label{toto20}
 \varepsilon_{i j}=\frac{9}{2}\delta_{i j}a^{3}
+\frac{e^2 m}{16\hbar^2}(\theta^{2}\delta_{i
j}-\theta_{i}\theta_{j})
\end{equation}

From this equation we make the following remarks, first
noncommutative polarizability is not proportional to $\delta_{ij}$
it has a more complicated tensorial structure. In other words
noncommutative Stark effect is direction dependent. It is due to
the fact that electron (in general any charged particle) in
noncommutative space has also an electric dipole moment. The
second remark is that $\varepsilon_{i j}^{(NC)}$ has a zero
eigenvalue ($ \varepsilon_{i j}\theta_{j}=0$). Physically it means
that noncommutative Stark effect vanishes if the external electric
field is parallel to $\vec{\theta}$. This is explicitly seen from
equation (\ref{toto16}). The third remark we make is of
observational significance; although the present experimental data
on Stark effect are not updated and  precise enough to improve the
current bounds on noncommutativity coming from other experiment
(chaichian et al (2001)). One may still study observational
prospects of noncommutative Stark effect. It is convenient to
present our results as the the ratio of noncommutative  to
commutative result,i.e
\begin{eqnarray}
\label{toto21}
\Delta_{NC}&=&\frac{E^{(NC)}}{E^{C}}\nonumber\\
&=&\frac{\frac{e^2
 m}{32\hbar^2}|\vec{E}\times\vec{\theta}|^2}{\frac{9 a^{3}
 }{4}|\vec{E}|^2}=\frac{|\vec{\theta}|^2}{72a^4}sin^2\phi,
\end{eqnarray}
where $\phi$ is the angle between $\vec{\theta}$ and $\vec{E}$. It
is worth noting that the above ratio is independent of
$\alpha_{QED}$. If the precision of the Stark effect is
$P_{Stark}$, the ratio $\Delta_{NC}$ should be within the
error-bars of the experiment or $\Delta_{NC}\preceq P_{Stark}$.
From this one can infer an upper- bound on $\theta$, or
equivalently a lower bound on the noncommutativity scale.

\section{Conclusion}
In this paper we have shown that there is a contribution to the
quadratic Stark effect due to the noncommutativity of the
coordinates. We have obtained the correction to the ground state
energy using two methods and by combining the two we have derived
the sum rule of the mean oscillator strength. If the external
electric field $\vec{E}$ is parallel to $\vec{\theta}$ then there
will be no shift in the ground state energy of the the H-atom,
therefore in this case we have only the ordinary quadratic Stark
effect. If the the two vectors are not parallel then the
polarization of the ground state is no longer diagonal. From which
we may infer some lower bound on the noncommutativity scale.


\vspace*{7mm}

{\bf Acknowledgments:}

We would like to thank M.M. Sheikh-Jabbari for useful discussions
and suggestions. N.C. would like to thank the Abdus Salam ICTP,
and SISSA for support and hospitality throughout these years.


\newpage

\bibliographystyle{phaip}

\end{document}